\documentclass[review]{elsarticle}

\usepackage{lineno,hyperref}
\usepackage{amsmath}
\usepackage{amssymb}
\usepackage{mathdots}
\usepackage{graphicx}
\usepackage{algorithmic}
\usepackage{textcomp}
\usepackage{graphicx}
\usepackage{caption}
\usepackage{subcaption}
\usepackage{cuted}
\usepackage{siunitx}

\DeclareSIUnit{\count}{count}

\newtheorem{Remark}{Remark}

\newtheorem{Theorem}{Theorem}

\journal{arXiv}
	
\begin{document}

\begin{frontmatter}

\title{A General Controller Scheme for Stabilization \& Disturbance Rejection with Application to Non-Linear Systems and its Implementation on 2 DOF Helicopter}

\author{Justin Jacob\corref{mycorrespondingauthor}}
\cortext[mycorrespondingauthor]{Corresponding author}
\ead{justinjacob@iitb.ac.in}
\author{Navin Khaneja}
\ead{navinkhaneja@gmail.com}
\address{Systems and Control Engineering Department, \\Indian Institute of Technology, Bombay}

\begin{abstract}
A general controller scheme for stabilizing a non-linear system, which has its origin from the linear system theory, is proposed in this paper. The proposed controller can stabilize the non-linear system subjected to initial conditions.  An effective way to obtain the controller parameters is presented with the knowledge of the system model. The controller is designed for the linear time-invariant (LTI) system, which can reject any disturbance acting on it. Paper emphasis the idea of an integrator controller in disturbance rejection. The concept is extended to the application to non-linear systems where the non-linearities are assessed as the disturbance to the refined linear part of the system. Boundedness and convergence of the non-linear system with the controller are proved to justify system stabilization. Hardware implementation of the controller on the 2 dof helicopter model is presented with experimental results, which validates the proposed control scheme.
\end{abstract}


\end{frontmatter}


\section{Introduction}
 
Controllers are an integral part of any industrial system. From simple to complex, a wide range of controllers can be found in the literature\cite{PUSHPKANT20161038}\cite{7364875}. Here a general controller scheme for a linear time-invariant (LTI) system, which can be applied to a non-linear system, is proposed. It's a model-based control scheme formulated based on the linear system theory. The parameter design of the controller is the simplest of all controllers as it is centered on the pole placement by state feedback.\cite{jacobpole}. The paper presents an effective way in finding the controller parameters, which is effortless and straightforward, assuming the system dynamics to be known. The controller parameters are designed based on the refined linear part of the system model with the knowledge of the system characteristic coefficients and the desired characteristic coefficients\cite{chen1999linear}. The performance of a second-order system can be easily related to the characteristic coefficients. By dominant pole analysis\cite{dominantpole}, the model can be extended and approximated to any degree.  \\
Knowledge of the system alone doesn't guarantee a model-based controller to stabilize the plant. Unknown disturbance and parameter variations cause changes to the desired system response. Keyser discusses disturbance modelling and its importance in a model-based controller\cite{1223451}, and Davison proposes a productive way for pole placement to stabilize the linear system with constant disturbance\cite{DAVISON}. This paper presents an approximation to disturbance acting on the system and the importance of an integrator as a disturbance rejection controller. The controller also stabilizes the system for any initial conditions.  \\
Most of the mechatronic systems have non-linear second-order coupled dynamics. So this paper analyzes the application of the proposed controller on a non-linear 2 dof helicopter model, which resembles a general system. One could find a wide range of control strategies in the literature for the 3 dof helicopter, and Kocagil\cite{kocagil2017controller} presents a review of it. The non-linear part of the dynamics is considered as disturbance while modelling. Stabilization of the system with the help of an integrator when the states are within the region of interest is shown. The closeness of the model to a linear system depends on the amount and types of non-linearities\cite{khalil2002nonlinear}. Usually, sinusoidal or exponential terms contribute to these non-linearities, and these can be approximated to the desired order by the Taylor series expansion. Typically, the third and higher degree terms contribute significantly less when compared to the fundamental and second-degree terms\cite{whittaker_taylor_cauchy}. For this reason, here, the non-linearities are approximated till the second-degree terms. Hardware implementation of the proposed control law on the 2 dof helicopter model is presented in this paper. The experiment results validate the application of the controller scheme on non-linear systems for control and stabilization.\\

\section{Theory}

Differential equations of a coupled system are combined to form a single equation; hence a general $n^{th}$ order linear time-invariant (LTI) system can be described by,

\begin{equation}\label{generalsystem}
\dfrac{d^n x}{dt^n} + \sum_{i=1}^{n} a_i \dfrac{d^{(i-1)} x}{dt^i} = u + T
\end{equation}
here the simplest of cases with a single input, $u \in \mathbb{R}$, and a single output which is the state, $x \in \mathbb{R}$, and other states as the successive derivatives are taken. $T \in \mathbb{R}$ is the disturbance acting on the system, with $\| T \| < \infty $.

\begin{Theorem} \label{objective}

Any system of the form Eq.\eqref{generalsystem} can be stabilized using the control input  \\
\begin{equation}\label{controller}
u = b_0 \int_0^t y \, dt + \sum_{i=1}^{n} b_i \dfrac{d^{(i-1)} y}{dt^{i-1}}.
\end{equation}
where $y = x_d - x$ with $x_d \in \mathbb{R}$ as the desired output and $b_i \in \mathbb{R}$ as the gain constants corresponding to the states. 

\end{Theorem}

\begin{Remark}
\textbf{For a system of $n^{th}$ order, $(n~-~1)$ derivatives, a proportional and an integral controller part are necessary for the control law to control, stabilize, and reject disturbance. }
\end{Remark}

\subsection{Proof}

\subsubsection{Disturbance Rejection}

\begin{figure}[!h]
\centering
\includegraphics[scale=.9]{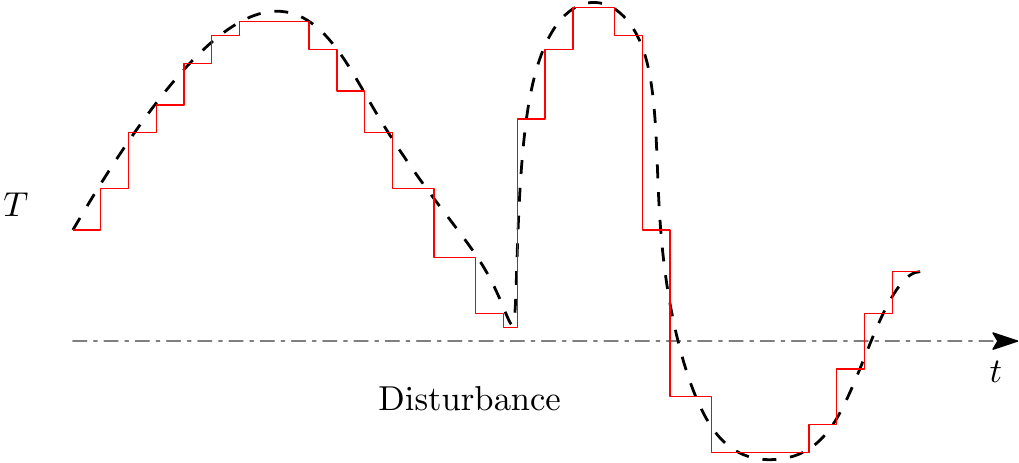}
\caption{Piece-wise constant disturbance.}
\label{disturbance}
\end{figure}

Assuming the system to be stable and the disturbance is approximated to piecewise constant. The disturbance is now a series of step inputs, as in figure~\ref{disturbance}, where a random disturbance is examined.

Now the objective is to eliminate the response of the system to these constant input disturbances by feedback. Adding the $\frac{1}{s}$ block to the system, the input to the system becomes impulse input, and the output becomes impulse response (IR).

\begin{figure}[!h]
\centering
\includegraphics[scale=1]{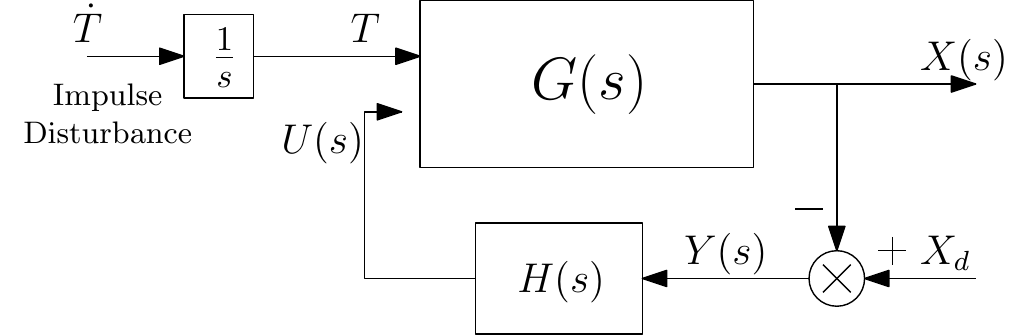}
\caption{Model of the effective system.}
\label{model}
\end{figure} 
The output of the effective system is obtained as

\begin{equation} \label{disturbance_op}
 X(s) = \frac{1}{s} \left(\frac{G(s)}{1 + G(s) H(s)} \right) + \frac{x_d}{s} \left(\frac{G(s) H(s)}{1 + G(s) H(s)} \right).
\end{equation}
$H(s)$ is chosen to stabilize the system from the disturbance, which is given by the first part in Eq.\eqref{disturbance_op}. The main intention is to eliminate the existence of $\frac{1}{s}$ from the output equation. The pole at the origin makes the system marginally stable; hence any disturbance to the system may cause instability to the system. With $H(s) = \frac{b_0}{s} $, which in turn tells $u = b_0 \int_0^t y(\tau) d\tau$, an \textbf{integral controller}, eliminates $\frac{1}{s}$ term. The key idea is to represent the disturbance as a sequence of step inputs. For small-time $t = \epsilon$ one can obtain the sequence step functions which will resemble the disturbance. And these step responses will decay as the system considered is a stable one. The response to one of the step input is

\begin{equation}\label{finalvalue}
X(s) = \dfrac{1+ b_0 \, x_d/s}{s^{n+1} + a_n s^n + \cdots + a_1 s + b_0}.
\end{equation}  
Applying the final value theorem\cite{fvt_ivt} on Eq.\eqref{finalvalue}, $x$ $\longrightarrow x_d$ as $t$~$\longrightarrow~\infty$. Most of the practical disturbance have a span very much less than the operational period. Hence all it's effect, will be eliminated over time.  

\begin{Remark}\label{remark1}
\textbf{Theoretically, every disturbance can be modelled by a sequence of impulse input, and an integrator in the controller eliminates it.}\\
\end{Remark}

\subsubsection{Stabilization}

A stable system is not always guaranteed; hence the first assumption of stability does not hold in every case. So the system is made stable with the idea of pole placement by giving feedback. The key idea here is to manipulate each coefficient in the denominator of the system. 

\begin{equation}\label{stablecontroller}
X(s) = \dfrac{\dfrac{x_d}{s}\left(\dfrac{1}{s^n + \sum_{i=1}^{n} a_is^{i-1}} \right)H(s)}{1+ \left(\dfrac{1}{s^n + \sum_{i=1}^{n} a_is^{i-1}} \right) H(s)}
\end{equation}
$H(s)$ has to account for all the coefficients corresponding to $s^0$ to $s^{n-1}$. Thus, $H(s) = b_1 s^0 + b_2 s^1 + \cdots + b_n s^{n-1} $,  and the input $ u = b_1 y + b_2 \dfrac{dy}{dt} + b_3 \frac{d^2 y}{dt^2} + \cdots + b_{n} \dfrac{d^{n-1} y}{dt^{n-1}} $, which is a combination of \textbf{derivative controllers}. Taking proportional term as the zeroth derivative, this shows \textbf{a one-to-one relationship between the number of derivatives to the order of the system}. Combining the disturbance rejection controller and the stabilization controller the general controller scheme is obtained. Applying the general controller scheme to Eq.\eqref{stablecontroller}, the integrator part just shifts the coefficients to the next power and the same happens to the derivative controller part. In effect the denominator becomes

\begin{equation}
s^{n+1} + \sum_{i=1}^{n} (a_i+b_i)s^{i} + b_0 .
\end{equation}
An appropriate choice of gains will result in a stable system. When taking the Laplace transform with non zero initial conditions, an additional term in the numerator appears corresponding to the initial values. This term makes a proper rational, and its effect goes to zero as $t$~$\longrightarrow~\infty$.

\section{Modelling 2 DOF}

\begin{figure}[!h]
\centering \leftskip=-2cm
\includegraphics[scale=1]{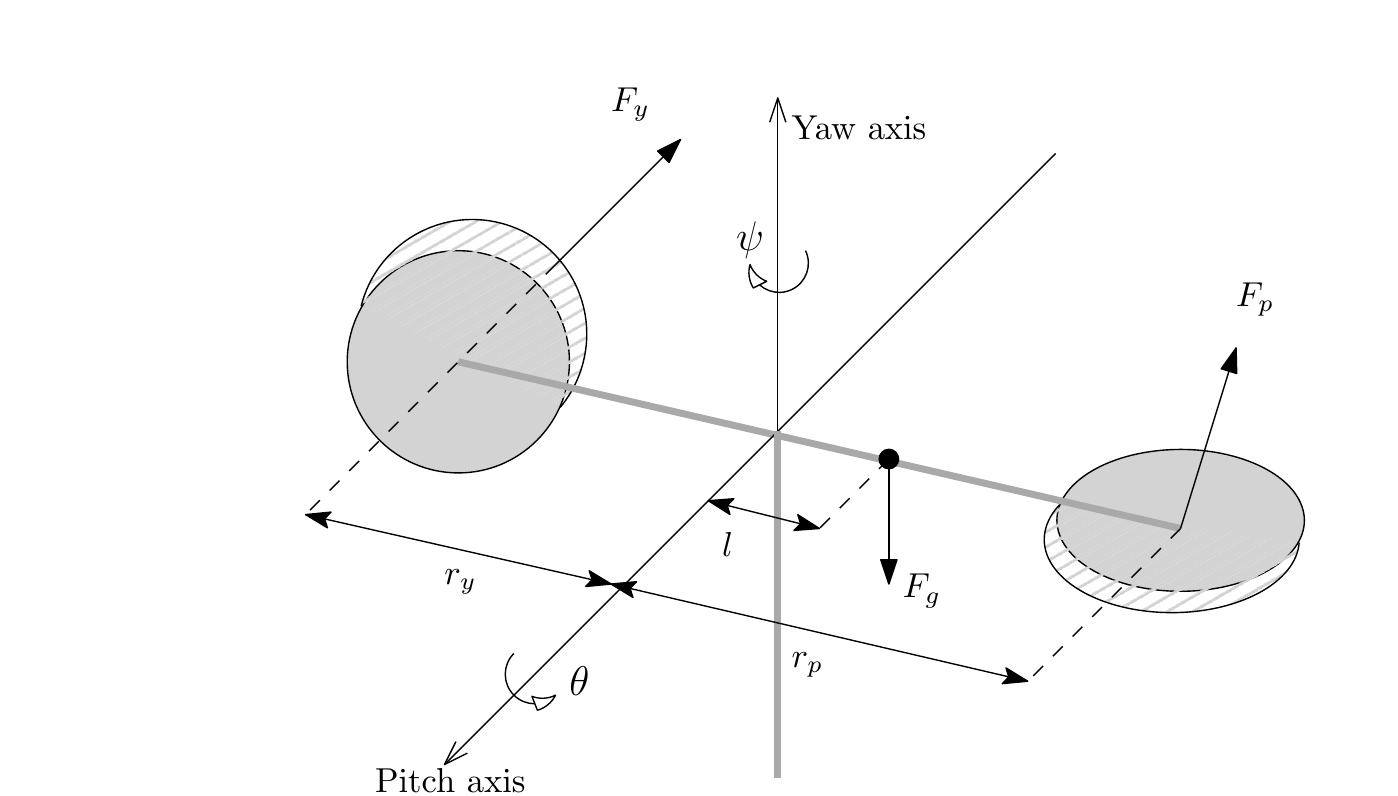}
\caption{Free body diagram.}
\label{FBD_2DOF}
\end{figure}

The free body diagram of the 2 dof helicopter is shown in figure \ref{FBD_2DOF}.

Here $\theta$ is the pitch angle, and $\psi$ is the yaw angle. $F_p$ and $F_y$ are the pitch and yaw thrust forces, respectively. Here the center of mass ($m$, moving mass of helicopter) is shifted more towards the nose end of the helicopter, and $l$ is the length from the hinge to it.\\

Euler-Lagrange's method\cite{morin2008introduction} is used to obtain the dynamic model of the system where the Lagrangian coordinates ($q$) are the pitch and yaw angles and torque ($T$) acting on pitch and yaw axis as the non conservative forces. From Euler Lagrange's equation $\dfrac{d}{dt}\dfrac{\partial L}{\partial \dot{q_i}} - \dfrac{\partial L}{\partial q_i} + \dfrac{\partial D}{\partial \dot{q_i}} = T_i$ where $L$ is the lagrangian of the system (Total energy = potential energy + kinetic energy), and $D$ is the Rayleigh dissipation function (viscous friction forces). $D = \dfrac{1}{2} B_p \dot{\theta^2} + \dfrac{1}{2} B_y \dot{\psi^2}$ where $B_p$ and $B_y$ are the pitch and yaw viscous friction thrust coefficients, respectively. 

\subsection{Equation of Motion}

Total potential energy is
\begin{equation}
PE = m g l \sin\theta .
\end{equation}
The total kinetic energy includes the kinetic energy due to rotation and kinetic energy due to translational motion. 

\begin{equation}
KE_1 = \dfrac{1}{2} J_p \dot{\theta}^2 + \dfrac{1}{2} J_y \dot{\psi}^2
\end{equation}
\begin{equation}
KE_2 = \dfrac{1}{2} m [ ( l \dot{\theta} )^2 + ( l \cos\theta \dot{\psi} )^2 ]   
\end{equation}
here equations are obtained assuming the pitch angle to change first and later the yaw. The Lagrangian of the system is $ KE_1 + KE_2 - PE $, as the potential energy is acting against the direction of motion. The Euler Lagrange's equation corresponding to pitch and yaw coordinates gives us the equation of motion of the helicopter,
\begin{equation}\label{pitchgeneraleq}
J_p \ddot{\theta} + B_p \dot{\theta} = T_{\theta} - m g l \cos\theta - m l^2 \cos\theta \sin\theta \dot{\psi}^2 - m l^2\ddot{\theta}
\end{equation}
\begin{equation}\label{yawgeneraleq}
J_y \ddot{\psi} + B_y \dot{\psi} = T_{\psi} + 2m l^2 \cos\theta \sin\theta \dot{\theta} \dot{\psi} - m l^2 \ddot{\psi} \cos^2\theta .
\end{equation}

\subsection{Small Angle Model}

Approximating the trigonometric relations by the Taylor series\cite{whittaker_taylor_cauchy}, the non-linear terms in the equation are reduced. With second-order approximation, the equation of motion becomes

\begin{equation}\label{pitchsmalleq}
( J_p +  m l^2 )\ddot{\theta} + B_p \dot{\theta} = T_{\theta} - m g l +  \dfrac{m g l \theta^2}{2} - m l^2 \dot{\psi}^2 + \dfrac{m l^2 \theta^2 \dot{\psi}^2}{2}
\end{equation}
\begin{equation}\label{yawsmalleq}
( J_y + m l^2 )\ddot{\psi} + B_y \dot{\psi} = T_{\psi} + 2 m l^2\left(1 - \dfrac{\theta^2}{2} \right) \dot{\theta} \dot{\psi} + m l^2 \theta^2 \ddot{\psi} .
\end{equation}
The next sections exhibit the stabilizability of a non-linear system by the general controller scheme with the help of the 2 dof helicopter non-linear model.

\section{Application on Non-Linear System}

\subsection{Stability Analysis}

Since the order of the system in Eq.\eqref{pitchsmalleq} and Eq.\eqref{yawsmalleq} remains second order; a proportional, derivative, and an integral control actions for each is required. Let the states be $z_1 = \int \theta -\theta_d$ and $z_2 = \int \psi - \psi_d$ and its first and second derivatives, where $\theta_d$, $\psi_d$ are the desired pitch and yaw angles. The control inputs required corresponding the general controller scheme are; $T_{\theta} = -k_1 z_1 - k_2 z_3 - k_3 z_5$ and $T_{\psi} = -k_4 z_2 - k_5 z_4 - k_6 z_6$, where $k_i$ are the gain constants. Neglecting $ml^2 \theta^2$ coming with $\ddot{\psi}$ term to bring more components to the linear part, which forms the refined linear part, and rearranging it to the state space form gives

\begin{equation}\label{stateNL}
\begin{split}
\begin{bmatrix}
\dot{z_1}\\ \dot{z_2}\\ \dot{z_3}\\ \dot{z_4}\\ \dot{z_5}\\ \dot{z_6}
\end{bmatrix}
=
\begin{bmatrix}
 0 & 0 & 1 & 0 & 0 & 0 \\
 0 & 0 & 0 & 1 & 0 & 0 \\ 
 0 & 0 & 0 & 0 & 1 & 0 \\
 0 & 0 & 0 & 0 & 0 & 1 \\
 -k_1 & 0 & -k_2 + \alpha_1 m g l \theta_d & 0 & -k_3 - \alpha_1 B_p & 0 \\
 0 & -k_4 & 0 & -k_5 & 0 & -k_6 - \alpha_2 B_y
\end{bmatrix}
\begin{bmatrix}
z_1\\ z_2\\ z_3\\ z_4\\ z_5\\ z_6 
\end{bmatrix}
+\\
\begin{bmatrix}
0 \\ 0 \\ 0 \\ 0 \\ -\alpha_1 mgl +\dfrac{a_1mgl \theta_d^2}{2} \\ 0
\end{bmatrix}
+
\begin{bmatrix}
0 \\ 0 \\ 0 \\ 0 \\ \dfrac{\alpha_1 m g l z_3^2}{2} - \alpha_1 m l^2 z_6^2 + \alpha_1 m l^2 \theta_d^2 z_6^2 + 2\alpha_1 m l^2 \theta_d z_3 z_6 + \alpha_1 m l^2 z_3^2 z_6^2 \\ \alpha_2 m l^2(2 + \theta_d^2)z_5 z_6 + 2 \alpha_2 m l^2 \theta_d z_3 z_5 z_6 + \alpha_2 m l^2 z_3^2 z_5 z_6
\end{bmatrix}
\end{split}
\end{equation}
where $\dfrac{1}{( J_p +  m l^2 )} = \alpha_1$ and $\dfrac{1}{( J_y +  m l^2 )} = \alpha_2$. Eq.\eqref{stateNL} is of the form $\dot{Z} = A Z + C + N$ and $A$ is the refined state matrix. With proper gain values of $T_{\theta}$ and $T_{\psi}$, one can stabilize the equilibrium point, and the constant term can be compensated by adding a bias to the control input. The solution for this system now becomes 
\begin{equation}\label{statesolution}
Z(t) = e^{At}Z(0) + \int_0^t e^{A(t-\tau)} N(z,\tau) d\tau
\end{equation}

\subsection{Boundedness}

Taking the euclidean norm and applying triangular inequality on Eq.\eqref{statesolution}, 
\begin{equation}\label{eqbd1}
\|Z(t)\| \leq \|e^{A t} Z(0)\| + \| \int\limits_{0}^{t}e^{A(t-\tau)}\,N(z,\tau)\,d\tau\|
\end{equation}
diagonalising $A$ by a similarity transformation $M \Sigma M^{-1}$, where $M$ is the model matrix whose columns are the eigenvectors. Then
\begin{equation}
e^{A} = M e^{\Sigma} M^{-1}
\end{equation} 
taking the norm 
\begin{equation}\label{diagonal}
\| e^{A} \| \leq \|M\| \|e^{\Sigma}\| \|M^{-1}\|
\end{equation}
where $\Sigma = diag(\lambda_1, \lambda_2, \lambda_3, \lambda_4, \lambda_5, \lambda_6)$ and $\lambda_i \in \mathbb{C}$ are the eigenvalues of matrix A.

\begin{Remark}\label{remark2}
\textbf{If the matrix $M$ is orthogonal (unitary), then 2-norm and the inner product are invariant under multiplication by it.}\\
\end{Remark}
In general let $\|M\| \|M^{-1}\| = \beta$, and as all the eigenvalues are negative, $e^{\lambda}$ can at most attain 1. Hence
\begin{equation}
\| e^{A} Z(0) \| \leq \beta \| Z(0) \|.
\end{equation}
Now for small $\| Z \|$, $\| z_i \| \leq \| Z \|$ and
\begin{equation}
\| N(z,t) \| \leq \kappa\| Z(t) \|^2 
\end{equation} 
where $\kappa^2 = (\frac{\alpha_1mgl}{2} + 2\alpha_1ml^2 \theta_d + \alpha_1ml^2 \theta_d^2 )^2 + (3 \alpha_2ml^2 + 2 \alpha_2ml^2 \theta_d + \alpha_2ml^2 \theta_d^2 )^2 $. Let $\|Z(t)\|$ is bounded by constant $\gamma$, then 
\begin{align}\label{integralbound}
\| \int\limits_{0}^{t} e^{A(t-\tau)} \, N(z,\tau)  \,d\tau \| &\leq \beta \, \kappa\, \gamma^2 \| \int\limits_{0}^{t} e^{\lambda_1(t-\tau)} \, d\tau \| \\ &\leq \beta \, \kappa\, \gamma^2 \| \dfrac{e^{\lambda_1} - 1}{\lambda_1} \| \\ &\leq \dfrac{\beta \, \kappa\, \gamma^2}{|\lambda_1|} .
\end{align}
Combining both terms and substituting back in Eq.\eqref{eqbd1}
\begin{equation}
\| Z(t)\| \leq \beta \| Z(0) \| + \dfrac{\beta \, \kappa\, \gamma^2}{|\lambda_1|} \leq \gamma .
\end{equation}
	
\begin{figure}[!h]
\begin{center}
\includegraphics[scale=.9]{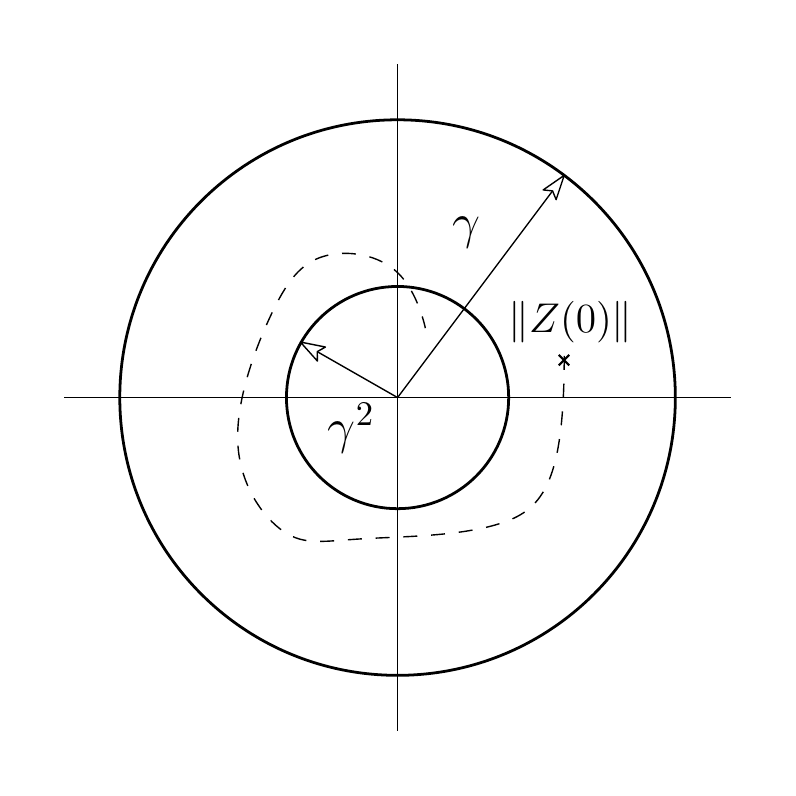}
\caption{Illustration of bounds on states and non-linearity.}
\label{2dof_bounds}
\end{center}
\end{figure}
Hence for small initial conditions, $\gamma^2$ will be less than $\gamma$, and the states never get out of the bound.

\subsection{Convergence}
Let $t_1 = t_2 + t_0$, and $t_0$, $t_1$, $t_2$ $\in \mathbb{N}$, then 
\begin{equation} 
\begin{split}
Z(t_1)=e^{A(t_2)}e^{A(t_0)}\,Z(0) + \int\limits_{0}^{t_2}e^{A(t_2-\tau)}e^{A(t_0)}\,N(z,\tau)\,d\tau \\+ \int\limits_{t_2}^{t_2+t_0}e^{A(t_2-\tau)}e^{A(t_0)}\,N(z,\tau)\,d\tau .
\end{split} 
\end{equation}
The difference between two time instance
\begin{equation} 
\begin{split}
Z(t_1) - Z(t_2) = e^{A(t_2)} \left(e^{A(t_0)} - I\right)\,Z(0) \\+ \int\limits_{0}^{t_2}e^{A(t_2-\tau)} \left(e^{A(t_0)} - I\right)\,N(z,\tau)\,d\tau \\+ \int\limits_{t_2}^{t_2+t_0}e^{A(t_2-\tau)}e^{A(t_0)}\,N(z,\tau)\,d\tau .
\end{split}
\end{equation}
Taking the norm and substituting the upper bounds for non-linear term,
\begin{equation} \label{converge}
\begin{split}
\|Z(t_1) - Z(t_2)\| \leq \|e^{A(t_2)} \left(e^{A(t_0)} - I\right)\| \, \|Z(0)\| \\+ \int\limits_{0}^{t_2} \|e^{A(t_2-\tau)} \left(e^{A(t_0)} - I\right)\| \,\gamma^2 d\tau \\+ \int\limits_{t_2}^{t_2+t_0} \|e^{A(t_2-\tau)}e^{A(t_0)}\| \, \gamma^2 d\tau .
\end{split}
\end{equation}
Note that all the terms except terms with $t_2$ are constants, and $e^{A}$ is bounded by the exponential of the maximum of eigenvalue. Evaluating the integral
\begin{equation} \label{conergebound}
\| \int\limits_{0}^{t_2} e^{A(t_2-\tau)} d\tau \| \leq \| \int\limits_{0}^{t_2} e^{\lambda_{max}(t_2-\tau)} d\tau \|  \leq \| \dfrac{e^{\lambda_{max}t_2} -1}{\lambda_{max}} \|
\end{equation}
as all the eigenvalues are negative, its a finite value, hence
\begin{equation}
\|Z(t_1) - Z(t_2)\| \leq \varepsilon .
\end{equation}
The integral term in Eq.\eqref{statesolution} is bounded by $\dfrac{1}{\lambda_{max}}$, and the rate of decay of $1^{st}$ term to zero and the rising of the $2^{nd}$ to the upper bound is same. Hence $X(t)$ always reduces if the initial states are small enough. This in turns reduce the non-linear term and from Eq.\eqref{converge}, as $t_1$,~$t_2$~$\longrightarrow~\infty$, RHS~$\longrightarrow~0 $. This resembles a Cauchy sequence\cite{whittaker_taylor_cauchy}, and the states converge over time. Boundedness along with convergence proves the system is stabilizable with the controller. This analysis can be generalized to any non-linear system.

\begin{Remark}\label{remark3}
\textbf{The bounded non-linearities are similar to the unpredicted disturbance acting on the refined linear model.}\\
\end{Remark}

\section{Experiment}

The set-up used in the laboratory is the Quanser 2 dof helicopter, which is mounted on a fixed base with two propellers (pitch and yaw) and is driven by DC motors. High-resolution encoders present in the fixed base measure the pitch and yaw angles, and are fed via the data acquisition board (DAB) to the computer. The DAB drives the actuators through a power amplifier. The main parameters associated with Quanser 2 dof Helicopter and the relation between torque and voltage are available in \cite{2dofdatas}. 

\begin{figure*}[!h]	
	\advance \leftskip-1cm
	\begin{center}
	\includegraphics[scale=.45]{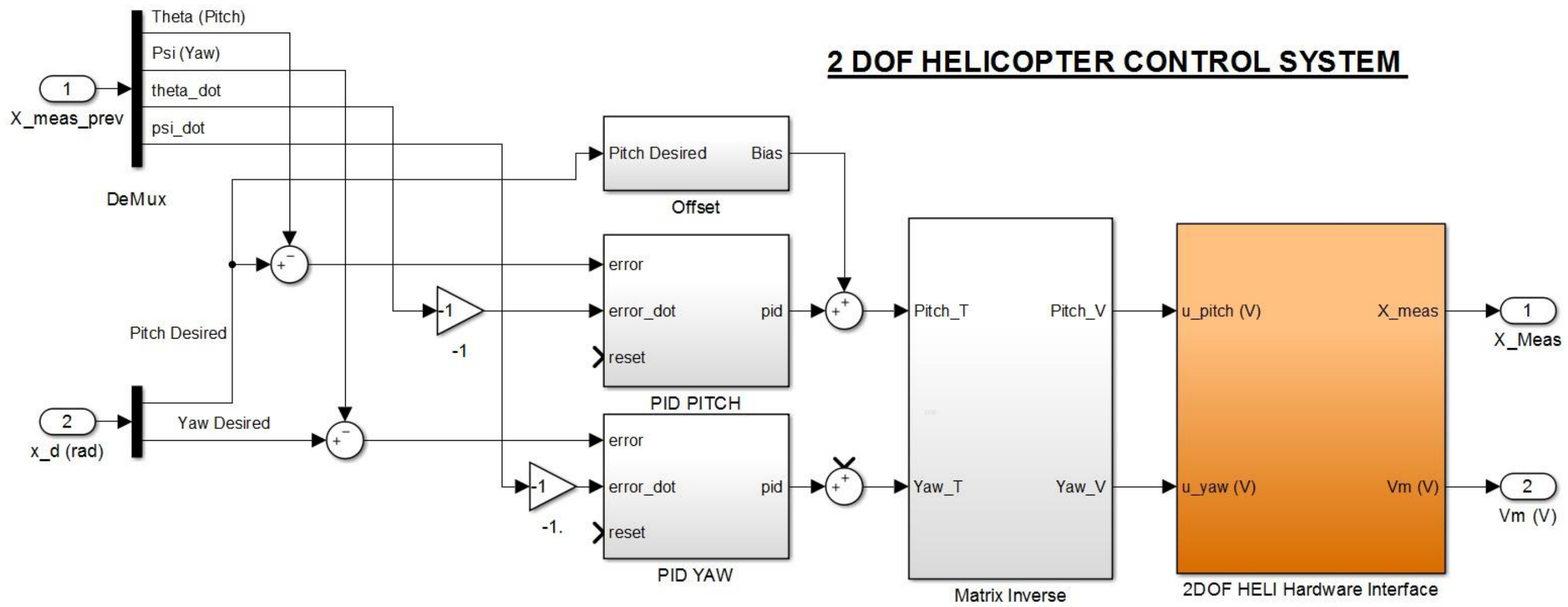}
	\caption{PID controllers with 2 dof helicopter system.}
	\label{2dof_controller_block}
	\end{center}
\end{figure*}

\begin{figure*}[!h]
\advance\leftskip-1cm
\begin{center}
\includegraphics[scale=.45]{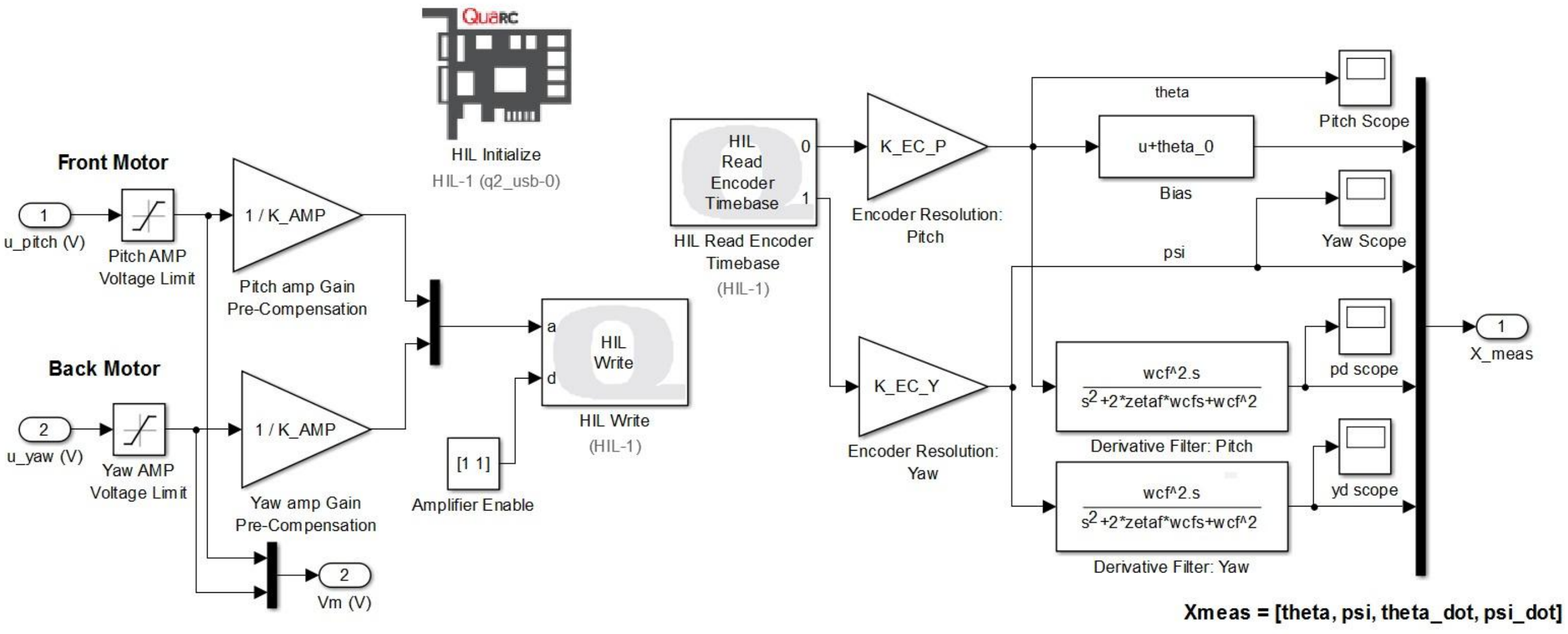}
\caption{Hardware data management block.}
\label{2dof_hardware}
\end{center}
\end{figure*}

Controller gain values obtained are from dominant pole analysis along with the parameters of a second-order system. Considering a peak overshoot of $1\%$ and settling time of $\SI{4}{\second}$, the gain values obtained are \\
\textbf{Pitch Controller Gains} :\\ $k_1 = 1.7431$, $k_2 = 2.4095$, $k_3 = 0.3849$ \newline
\textbf{Yaw Controller Gains} :\\ $k_4 = 1.8398$, $k_5 = 2.5431$, $k_6 = 0.9326$ \\
Separate performance characteristics can be utilized for the pitch and the yaw dynamics, and its the choice of the designer.
Interfacing between the hardware and software (MATLAB) is done with the help of the QUARC library. The input to the pitch motor is limited to $\pm \SI{24}{\volt}$ and for the yaw motor to $\pm \SI{15}{\volt}$. Pitch axis encoder resolution is set to $ 2\pi / (4*1024)\, \text{rad/count}$ and yaw to $2\pi /(8*1024)\, \text{rad/count}$. The initial position of the helicopter is taken corresponding to $\theta_0 = \SI{-40.5}{\degree}$ and $\psi_0 = \SI{0}{\degree}$. The upper limit of $\theta$ is $\SI{35}{\degree}$ due to the hardware setup. To obtain the derivatives of $\theta$ and $\psi$, a second-order low pass filter with a damping ratio of $.85$ and cut-off frequency of $40 \pi \, \text{Hz}$ is used. A back-calculation anti integral windup\cite{antiwindup} with an integral reset time of $\SI{1}{\second}$ is used along with the integral controller. Experiments were conducted on the 2 dof helicopter model with the proposed controller, which turned out to be a PID controller due to the second-order system dynamics.

\subsection{Results}

\begin{figure}[!h]
\includegraphics[scale=.3]{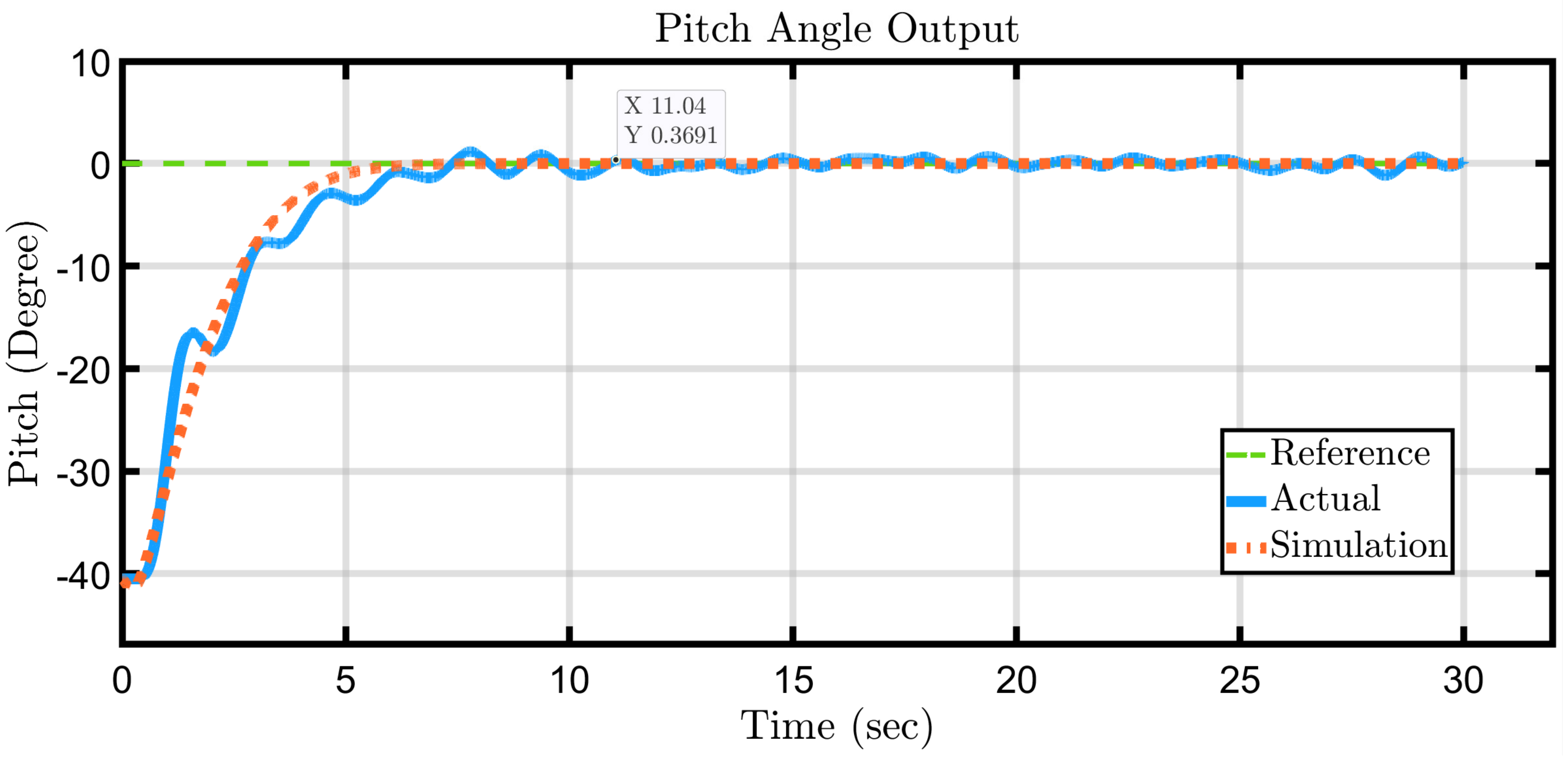}
\caption{Comparing pitch simulation and experimental output for $\theta_d = \SI{0}{\degree}$.}
\label{2dof_result21}
\end{figure}

\begin{figure}[!h]
\includegraphics[scale=.3]{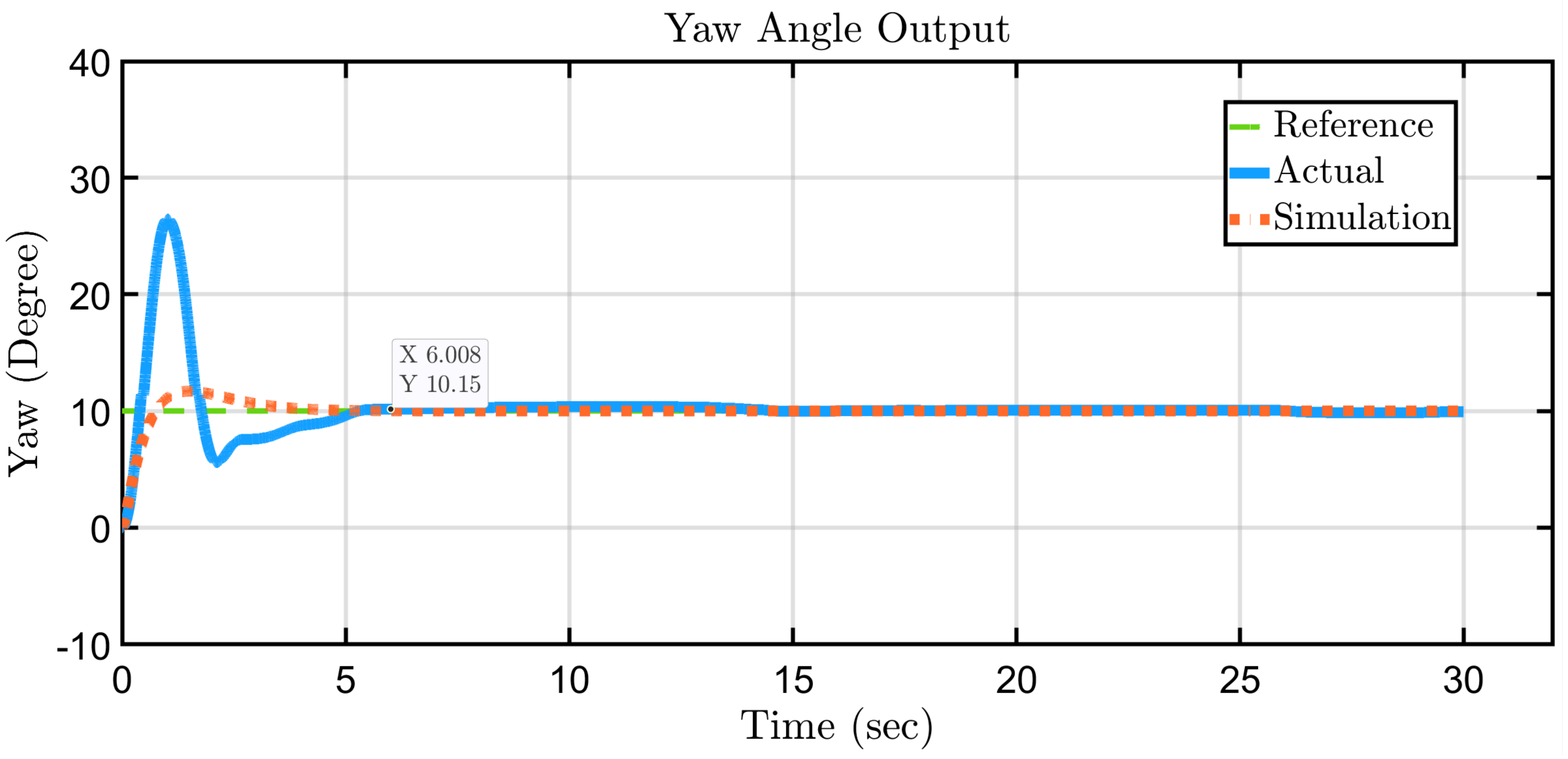}
\caption{Comparing yaw simulation and experimental output for $\psi_d = \SI{10}{\degree}$.}
\label{2dof_result22}
\end{figure}

\begin{figure}[!h]
\includegraphics[scale=.3]{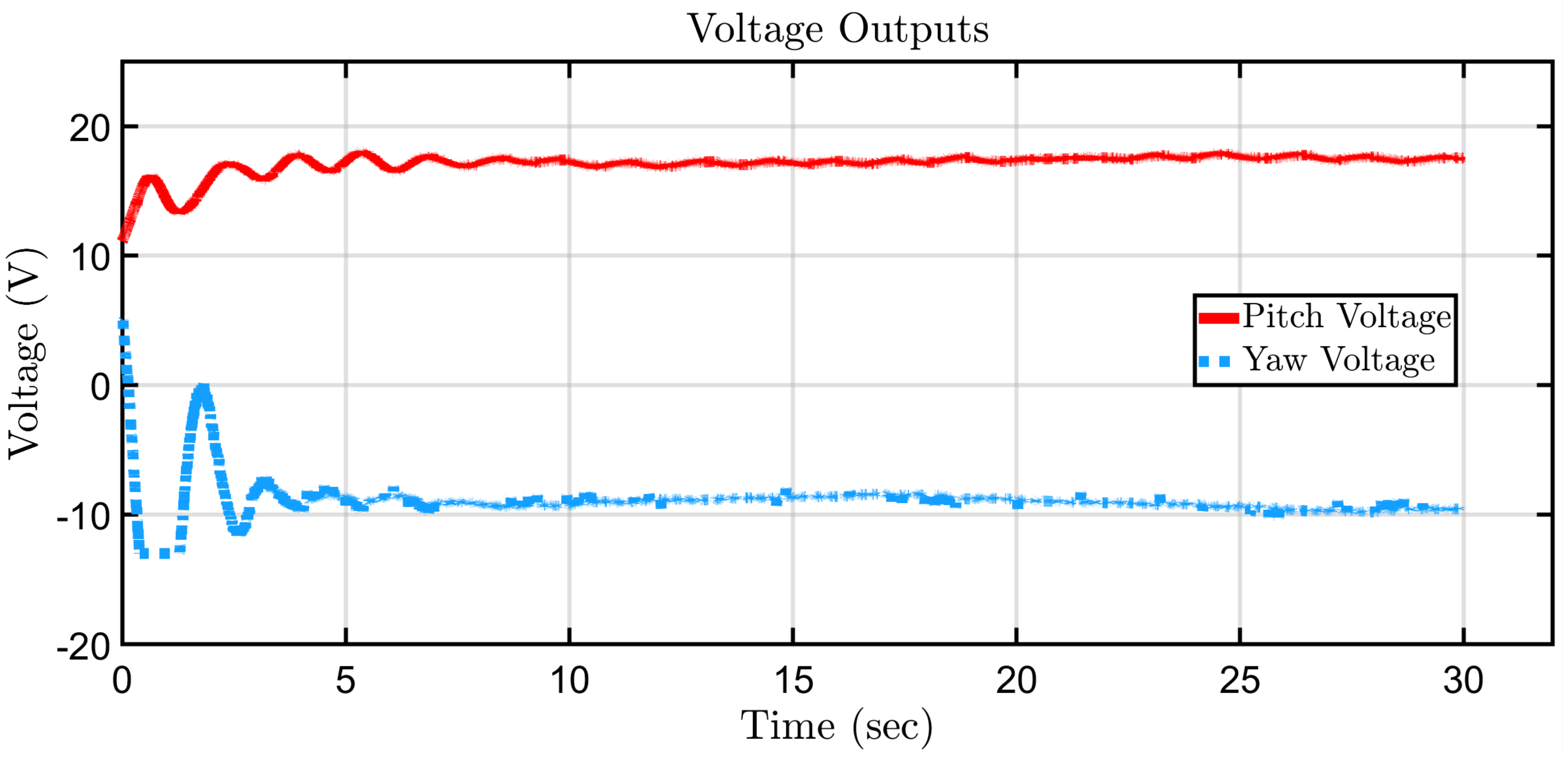}
\caption{Voltage outputs for pitch and yaw motors for $\theta_d = \SI{0}{\degree}$ and $\psi_d = \SI{10}{\degree}$.}
\label{2dof_result23}
\end{figure}

The experiment was conducted for $\SI{30}{\second}$ with the desired pitch of $0$ $\deg$ and desired yaw of $10$ $\deg$. Figure \ref{2dof_result21} shows the reference, actual and simulated output of the pitch angle in degrees. Figure \ref{2dof_result22} shows the reference, actual and simulated output of the yaw angle in degrees. The actual voltage output to the pitch and yaw motors, shown in figure \ref{2dof_result23}.

\subsection{Inference}

The system is highly non-linear and coupled; a slight overshoot in the simulation resulted in a more prominent oscillation in the experiment. Also, the system is very much prone to any external disturbance. So the gain values are designed such that the overshoot in simulation is very much negligible. If the set-point is far from the initial conditions, then the system takes a longer period to settle. Proper adjustment of the gain values can offset these conditions. Further, a low pass filter must accompany a higher-order derivative controller; else, it may damage the actuator. Optimization of control pulse based on different strategies, which change the controller parameters, can be implemented easily. The key idea is how the gains are related to the system dynamics and desired characteristic coefficient.

\section{Conclusion}

Modelling an accurate system is almost impossible, as there will be parameter variations and uncertainties acting on the system. A general controller scheme is proposed for a linear time-invariant (LTI) system, which takes care of all the uncertainties. The scheme is based only on the refined linear part of the model while keeping track of the non-linear function. The design of the controller parameters depends only on the refined linear system characteristic coefficients and the desired characteristic coefficients. The proposed controller is effortless and straightforward in design and can stabilize the non-linear system even though it's derived from the linearized model. The integrator interprets the non-linearities as a disturbance to the refined linear model. The proposed controller does a fine job in stabilizing the non-linear 2 dof helicopter model.

\bibliography{reference}

\begin{thebibliography}{10}
\expandafter\ifx\csname url\endcsname\relax
  \def\url#1{\texttt{#1}}\fi
\expandafter\ifx\csname urlprefix\endcsname\relax\def\urlprefix{URL }\fi
\expandafter\ifx\csname href\endcsname\relax
  \def\href#1#2{#2} \def\path#1{#1}\fi

\bibitem{PUSHPKANT20161038}
Pushpkant, S.~Jha,
  \href{https://www.sciencedirect.com/science/article/pii/S1877050916315538}{Comparative
  study of different classical and modern control techniques for the position
  control of sophisticated mechatronic system}, Procedia Computer Science 93
  (2016) 1038--1045, proceedings of the 6th International Conference on
  Advances in Computing and Communications.
\newblock \href {https://doi.org/https://doi.org/10.1016/j.procs.2016.07.307}
  {\path{doi:https://doi.org/10.1016/j.procs.2016.07.307}}.
\newline\urlprefix\url{https://www.sciencedirect.com/science/article/pii/S1877050916315538}

\bibitem{7364875}
P.~Faradja, G.~Qi, Robustness based comparison between a sliding mode
  controller and a model free controller with the approach of synchronization
  of nonlinear systems, in: 2015 15th International Conference on Control,
  Automation and Systems (ICCAS), 2015, pp. 36--40.
\newblock \href {https://doi.org/10.1109/ICCAS.2015.7364875}
  {\path{doi:10.1109/ICCAS.2015.7364875}}.

\bibitem{jacobpole}
J.~Jacob, S.~Das, N.~Khaneja, A concise method of pole placement to stabilize
  the linear time invariant mimo system, in: 2019 Sixth Indian Control
  Conference (ICC), IEEE, 2019, pp. 35--39.

\bibitem{chen1999linear}
C.-T. Chen, Linear system theory and design (1999).

\bibitem{dominantpole}
\href{https://www.sciencedirect.com/science/article/pii/B9780080334318500397}{Automatic
  tuning of pid controllers based on dominant pole design}, IFAC Proceedings
  Volumes 18~(15) (1985) 205--210, iFAC Workshop on Adaptive Control of
  Chemical Processes, Frankfurt a.M ,FRG, 21-22 October 1985.
\newblock \href
  {https://doi.org/https://doi.org/10.1016/B978-0-08-033431-8.50039-7}
  {\path{doi:https://doi.org/10.1016/B978-0-08-033431-8.50039-7}}.
\newline\urlprefix\url{https://www.sciencedirect.com/science/article/pii/B9780080334318500397}

\bibitem{1223451}
R.~De~Keyser, C.~Ionescu, The disturbance model in model based predictive
  control, in: Proceedings of 2003 IEEE Conference on Control Applications,
  2003. CCA 2003., Vol.~1, 2003, pp. 446--451 vol.1.
\newblock \href {https://doi.org/10.1109/CCA.2003.1223451}
  {\path{doi:10.1109/CCA.2003.1223451}}.

\bibitem{DAVISON}
E.~Davison, H.~Smith,
  \href{https://www.sciencedirect.com/science/article/pii/0005109871900999}{Pole
  assignment in linear time-invariant multivariable systems with constant
  disturbances}, Automatica 7~(4) (1971) 489--498.
\newblock \href {https://doi.org/https://doi.org/10.1016/0005-1098(71)90099-9}
  {\path{doi:https://doi.org/10.1016/0005-1098(71)90099-9}}.
\newline\urlprefix\url{https://www.sciencedirect.com/science/article/pii/0005109871900999}

\bibitem{kocagil2017controller}
B.~M. Kocagil, A.~{\c{C}}. Ar{\i}can, {\"U}.~M. G{\"u}zey, S.~{\"O}zcan, M.~U.
  Salamci, Controller designs for nonlinear systems with application to 3 dof
  helicopter model, Gazi University Journal of Science Part A: Engineering and
  Innovation 4~(3) (2017) 47--66.

\bibitem{khalil2002nonlinear}
H.~K. Khalil, Nonlinear systems, Prentice-Hall, 2002.

\bibitem{whittaker_taylor_cauchy}
E.~T. Whittaker, G.~N. Watson, A course of modern analysis, Courier Dover
  Publications, 2020.

\bibitem{fvt_ivt}
B.~Rasof,
  \href{https://www.sciencedirect.com/science/article/pii/0016003262909390}{The
  initial- and final-value theorems in laplace transform theory}, Journal of
  the Franklin Institute 274~(3) (1962) 165--177.
\newblock \href {https://doi.org/https://doi.org/10.1016/0016-0032(62)90939-0}
  {\path{doi:https://doi.org/10.1016/0016-0032(62)90939-0}}.
\newline\urlprefix\url{https://www.sciencedirect.com/science/article/pii/0016003262909390}

\bibitem{morin2008introduction}
D.~Morin, Introduction to classical mechanics: with problems and solutions,
  Cambridge University Press, 2008.

\bibitem{2dofdatas}
2 dof helicopter,
  \url{http:https://www.quanser.com/products/2-dof-helicopter/}, [Online;
  accessed 14-February-2020].

\bibitem{antiwindup}
H.~Markaroglu, M.~Guzelkaya, I.~Eksin, E.~Yesil, Tracking time adjustment in
  back calculation anti-windup scheme, 2006.
\newblock \href {https://doi.org/10.7148/2006-0613}
  {\path{doi:10.7148/2006-0613}}.

\end{thebibliography}
\bibliographystyle{elsarticle-num}

\end{document}